\documentclass[onecolumn,showpacs,floatfix,aps]{revtex4}
\usepackage{amssymb}
\usepackage[dvips]{graphics}
\newcommand{\bea}{\begin{array}}
\newcommand{\ear}{\end{array}}

\newcommand{\bege}{\begin{equation}}
\newcommand{\enge}{\end{equation}}

\newcommand{\beq}{\begin{eqnarray}}\newcommand{\benu}{\begin{enumerate}}\newcommand{\enu}{\end{enumerate}}
\newcommand{\eeq}{\end{eqnarray}}

\newcommand{\noi}{\noindent}

\begin{document}

\title{Anti-de Sitter curvature radius constrained by quasars in brane-world scenarios}

\author{R. da Rocha}
\email{roldao@ifi.unicamp.br}
\affiliation{IFGW, Universidade Estadual de Campinas,\\
CP 6165, 13083-970 Campinas, SP, Brazil.}

\author{C. H. Coimbra-Ara\'ujo}
\email{carlos@astro.iag.usp.br}
\affiliation{Departamento de Astronomia, Universidade de S\~ao Paulo, 05508-900
S\~ao Paulo, SP, Brazil.\\and\\
IFGW, Universidade Estadual de Campinas,\\
CP 6165, 13083-970 Campinas, SP, Brazil.}

\author{I. T. Pedron}
\email{thmar@vn.com.br}
\affiliation{Universidade Estadual do Oeste do Paran\'a, CP 91, 85960-000, Marechal C\^andido Rondon, PR, Brazil}


\begin{abstract}
This paper is intended to investigate the luminosity due to accretion of gas in supermassive black holes (SMBHs)
in the center of quasars, using a brane-world scenario naturally endowed with extra dimensions, whereon theories formulated
 introduce corrections in the field equations at high energies. SMBHs possess the necessary
highly energetic environment for the introduction of these corrections, which are shown to produce small deviations
 in all SMBH properties and, consequently, corrections in the accretion theory that supports quasars radiative processes.
 The radiative flux observed from quasars  indicates these deviations, from which the  magnitude of the AdS$_5$ bulk curvature radius, and consequently
the extra dimension compactification radius is estimated.
\end{abstract}

\maketitle

\section{Introduction}

Investigations concerning extra dimensions have been had considerable activity recently,
and have been motivated by possible extra dimensions arising within a TeV scale for quantum gravity.  The idea that the universe
is trapped on a membrane in some higher-dimensional spacetime may explain why gravity is so weak, and could be
 tested at high-energy particle accelerators. In this context, the observable universe can be described by
a $1+3$  manifold, the so-called \emph{brane}, embedded in a spacetime with $1+3+\delta$ dimensions, denominated \emph{bulk}, where $\delta$ is the number of extra
dimensions.  The idea of extra dimensions dates back to the 1890s, when Heinrich R. Hertz \cite{hertz} and Luther P. Eisenhart \cite{eise} dextrously proposed
the trajectories of a general conservative system in classical dynamics can be put into a one-to-one correspondence with the geodesics of a suitable Riemannian manifold in
one higher dimension \cite{barut1}. Thereafter in the 1920s,
 the second attempt to include extra dimensions  in a physical theory was made by Theodor Kaluza and Oskar Klein \cite{kaluza, Kaluza}.
 They suggested a unification of electromagnetism and gravity into a geometrical formulation,
involving one extra dimension that, when compactified, makes a unique formalism that unravels
the Lorentz and the U(1) gauge symmetry of electromagnetism simultaneously.

In spite of the possible existence of extra dimensions, they still remain up to now unaccessible to experiments.
Kaluza-Klein (KK) and string theories
indicate the introduction of compact extra dimensions \cite{5} of small radii of size of the order of the
Planck scale, $l_P \sim 10^{-33}$ cm. Those unaccessed dimensions can also be as large as millimeters, implying
deviations of the Newton's law of gravity at these scales, and the highest energy particle accelerators extend our range of sight to
include the weak and strong forces down to small scales, around 10-15 mm. Below about 1 mm, objects may be gravitating in more
dimensions. The electromagnetic, weak and strong forces, as well as all the matter
in the universe, would be trapped on a brane with three spatial dimensions. Only gravitons would be able to leave the surface
 and move throughout the full volume \cite{gr, zwi, zwi1, zwi2}.

At low energies, gravity is localized on the brane and general relativity is recovered, but at high energies,
 significant changes are introduced in gravitational dynamics, forcing general relativity to break down
to be overcome by a quantum gravity theory \cite{rov}.
A plausible reason for the gravitational force appear to be so weak can be its dilution in possibly existing extra dimensions
related to a bulk, where $p$-branes \cite{gr, zwi, zwi1, zwi2, bs, Townsend} are embedded. $p$-branes \cite{8, 17}
are good candidates for brane-worlds because they possess gauge symmetries \cite{zwi, zwi1, zwi2} and automatically incorporate a quantum theory of gravity. The
gauge symmetry arises from open strings, which can collide to form a closed string that can leak into the higher-dimensional
bulk. The simplest excitation modes of these closed strings
correspond precisely to gravitons \cite{qfts}.

It has been suggested the existence of infinite extra dimensions \cite{ken, Rubakov}, where particles in the bulk can be detectable in low
energy as 4-dimensional fields \cite{Dvali}. Also, an alternative scenario can be achieved using 5-dimensional theories over manifolds endowed with
the Randall-Sundrum  metric (RS), which induces a volcano barrier-shaped effective potential for gravitons around the brane \cite{Likken}.
The corresponding spectrum of gravitational perturbations has a massless bound state on the brane,
and a continuum of bulk modes with suppressed couplings to brane fields. These bulk modes introduce small
corrections at short distances, and the introduction of more compact dimensions does not affect the localization of matter fields.
However, true localization takes place only for massless fields \cite{Gregory}, and in the massive case the bound state becomes metastable,
being able to leak into the extra space. This is shown to be exactly the case for astrophysical massive objects,
where highly energetic stars and the process of gravitational collapse, which can originate black holes, leads to deviations from the $4D$ general relativity problem.

We can explore the idea that, if black holes and specially the supermassive black holes (SMBHs) present in the nucleus of galaxies and quasars,
 do cause deviations from the $4D$ general relativity,  these corrections should cause a small deviation in all SMBH properties. Consequently,
corrections in the accretion theory that supports quasars radiative processes. The radiative flux observed from quasars can then indicate
 these deviations and the AdS$_5$ bulk curvature radius.

In order to the article to be as self-contained as possible, in the first three Sections we make a brief review about
field equations on the brane. Although the description of black hole theory on the brane is relatively well known,
there are still some inconsistencies. One of them is the apparent discrepancy between the perturbation of the Schwarzschild form,
coming from the correction in the gravitational potential that arises from KK modes in RS formalism,
and the large bulk curvature radius, which is computed via quasar observations.

It is also shown an observational constraint between the given
perturbation in the Schwarzschild metric and the Schwarzschild
radius on the brane, showing the magnitude of the AdS$_5$ bulk
curvature radius, and subsequently the extra dimension
compactification radius. By a \emph{gedanken} experiment concerning
quasar luminosity, we obtain as a result a large bulk curvature
radius, and consequently a compactification radius $k \approx
10^{-14}$cm. This can indicate remarkable guidelines in the future
compactification scale investigations.

This article  is organized as follows: in Section 2,  Einstein field equations, written in terms of the Weyl tensor,
the tensor splittings on the brane and Codazzi equations, are used to derive a field equation on the brane. In Section 3 we derive a suitable
expression for this field equations, supposing the conservation of energy in the brane and the Israel-Darmois junction conditions.
In Section 4,  a Schwarzschild black hole on the brane is investigated in this context, and in Section 5
 the range of the AdS$_5$ curvature radii due to theory of accretion in quasars is computed, together with the extra dimension compactification radius.
Finally we present in Appendix some geometrical features
concerning the Weyl tensor, after proposing in the Concluding Remarks, two alternative  perspectives, respectively cohomological, concerning Penrose's
twistor theory \cite{pe1, pe2, ro1}, and geometrical, using brane-world
scenarios over non-orientable manifolds in order to circumvent all the
formal prerequisites necessary to approach the arising questions of the present formalism.

\section{Field equations on the brane}

In what follows, we adopt natural units, where $c=1$ and $G=1$, where $c$ denotes the light speed and $G$ the gravitational constant.
In this section we present and discuss the mathematical preliminaries necessary to explore the cosmological theory of black holes and
quasars in brane-world scenarios constraining the AdS$_5$ universe radius.
For a complete exposition about arbitrary manifolds and fiber bundles, see, e.g, \cite{frankel, naka, koni, moro, rowa}.
Hereon $\{e_\mu\}, \mu = 0,1,2,3$ [$\{e_A\}, A=0,1,2,3,4$] denotes a basis for the tangent space $T_xM$ at a point $x$ in $M$, where $M$ denotes the manifold
modelling a brane embedded in a bulk.
Naturally the cotangent space at $x$ has an orthonormal basis $\{\theta^\mu\}$ [$\{\theta^A\}$] such that $\theta^\mu(e_\nu) = \delta^\mu_\nu$.
If we choose a local coordinate chart, it is possible to represent $e_A = \partial/\partial x^A \equiv \partial_A$ and $\theta^A = dx^A$.
Given the extrinsic curvature $K = K_{AB}\theta^A \wedge\theta^B$, it is possible to project the bulk curvature on the brane.

Take $n = n^Ae_A$ a  vector orthogonal to $T_xM$ and let $y$ be the Gaussian coordinate orthogonal to the $T_xM$, on the brane at $x$, indicating
how much an observer upheave out the brane into the bulk.
In particular we have $n_AdX^A = dy$.
A vector $v = v^Ae_A$ in the  bulk  is split in components in the brane and orthogonal to the brane, respectively as
 $v = v^\mu e_\mu + ye_4 = (v^{\mu},y)$. Since the bulk is endowed with a
non-degenerate bilinear symmetric form $g$ that can be written in a coordinate basis, where $e^A:= dx^A,\; e_A = \partial/\partial x^A$, as
 $g = g_{AB}dx^A\otimes dx^B$,  the components of the metric on the brane and on the bulk are hereon denoted respectively by  $g_{AB}$ and
 $^{(5)}g_{AB}$, and related by \cite{mis}
\begin{equation}\label{neo}
{}^{(5)}g_{AB} = g_{AB} + n_An_B.
\end{equation}

The Gauss equation, relating curvatures in a 5-dimensional spacetime projected in a 4-dimensional one, can be written as \cite{frankel, mis, man}:
\begin{eqnarray}\label{duvida}
\hspace{-0.3cm}R^A_{\;\; BCD} &=& {}^{(5)}R^E _ {\;\; FGH}g_E^{\;\;A}g_{\;\;B}^Fg_{\;\;C}^Gg_{\;\;D}^H + K^A _ {\;\;C}K_{BD} - K^A _{\;\;D}K_{BC}\nonumber\\
&=&{}^{(5)}R^E _ {\;\; FGH}({}^{(5)}g_E^{\;\;A} - n_En^A)({}^{(5)}g_{\;\;B}^F - n^Fn_B)({}^{(5)}g_{\;\;C}^G-n^Gn_C)({}^{(5)}g_{\;\;D}^H -n_Dn^H) + K^A _ {\;\;C}K_{BD} - K^A _{\;\;D}K_{BC}
\end{eqnarray} Now it arises, from the antisymmetry of the first two and the last two indices of the Riemann tensor, that all but the terms
\begin{eqnarray}\label{uvida}
 {}^{(5)}R^E _ {\;\; FGH}{}^{(5)}g_E^{\;\;A}{}^{(5)}g_{\;\;B}^F{}^{(5)}g_{\;\;C}^G{}^{(5)}g_{\;\;D}^H + K^A _ {\;\;C}K_{BD} - K^A _{\;\;D}K_{BC}
\end{eqnarray}\noi survive in Eq.(\ref{duvida}) when we contract the Riemann tensor indices in order to obtain the Ricci scalar. We then consider, without loss of generality in what follows that   
\begin{eqnarray}\
R^A_{\;\; BCD}&=& {}^{(5)}R^A _ {\;\; BCD} + K^A _ {\;\; C}K_{BD} - K^A _ {\;\; D}K_{BC}.\nonumber
\end{eqnarray}
\noindent
Contracting the indices and left multiplying by $g^C _ {\;\; A}$, the components of the Ricci form in the bulk are derived
\begin{equation}\label{ricci}
R_{BD} = (^{(5)}g^C _ {\;\; A} - n^Cn_A)^{(5)}R^A _ {\;\;BCD} + ({\rm Tr}\; K)K_{BD} - K^C _ {\;\; D}K_{BC} + g^C _ {\;\; A}R^A _ {\;\; BCD}n^An^B,
\end{equation}
\noindent and, in the brane, we have:
\begin{equation}\label{ricci1}
R_{\mu\nu} = {}^{(5)}R_{BD}g_{\;\;\mu} ^{ B}g_{\;\;\nu} ^{ D} - {}^{(5)}R^A _{\;\; BCD}n^Cn_Ag_{\;\;\mu} ^{ B}g_{\;\;\nu} ^{ D} + ({\rm Tr}\; K) K_{\mu\nu} - K^C _{ \nu}K_{\mu C}.
\end{equation}
Using  eq.(\ref{ricci}), the Ricci scalar is calculated from the following equation:
\begin{eqnarray}
R  = g^{BD}R_{BD} &=& {}^{(5)}R - {}^{(5)}R_{BD}n^Bn^D - {}^{(5)}R^A_{\;\;BCD}n_An^C{}^{(5)}g^{BD} + ({\rm Tr}\; K)^2 - K^{BC}K_{BC}
+ R^A_{\;\;BCD}n_An^Bn^Cn^D
\end{eqnarray}
\noindent
Expressing  $g^{BD} =  g_{\;\;\mu}^Bg_{\;\;\nu}^Dg^{\mu\nu}$ yields
\begin{equation}\label{escalar}
Rg_{\mu\nu} = {}^{(5)}Rg_{BD}g_{\;\;\mu}^Bg_{\;\;\nu}^D - {}^{(5)}R_{BD}n^Bn^Dg_{\mu\nu} - {}^{(5)}R^A_{\;\;BCD}n_An^Cg_{\;\;\mu}^Bg_{\;\;\nu}^D +
 [({\rm Tr}\; K)^2 - ({\rm Tr}\; K^2)]g_{\mu\nu}.
\end{equation}
\noindent
From eqs.(\ref{ricci1}) and (\ref{escalar}), the tensor $G_{\mu\nu}:= R_{\mu\nu} - \frac{1}{2}Rg_{\mu\nu}$ is given by:
\begin{eqnarray}
G_{\mu\nu} &=& \left(^{(5)}R_{BD} -
\frac{1}{2}{}^{(5)}Rg_{BD}\right)g_{\;\;\mu}^Bg_{\;\;\nu}^D +
\frac{1}{2}{}^{(5)}R_{BD}n^Bn^Dg_{\mu\nu} +
\frac{1}{2}{}^{(5)}R^{\;\;A}_Cn_An^C g_{\mu\nu}
-{}^{(5)}R^A_{\;\;BCD}n_An^Cg_{\;\;\mu}^Bg_{\;\;\nu}^D \nonumber\\
&& - \frac{1}{2}{}^{(5)}R^A_{\;\;BCD}n^Bn^Cn^Dn_A g_{\mu\nu} + ({\rm Tr}\; K) K_{\mu\nu} - K^C _{ \nu}K_{\mu C}
 - \frac{1}{2}g_{\mu\nu}\left(({\rm Tr}\; K)^2 - ({\rm Tr}\; K^2)\right).
\end{eqnarray}
\noindent
Since $g_{BD} = {}^{(5)}g_{BD} - n_Bn_D$, then
\begin{eqnarray}
G_{\mu\nu} &=& \left(^{(5)}R_{BD} -
\frac{1}{2}{}^{(5)}R{}^{(5)}g_{BD}\right)g_{\;\;\mu}^Bg_{\;\;\nu}^D +
{}^{(5)}R_{BD}n^Bn^Dg_{\mu\nu} +
\frac{1}{2}{}^{(5)}R^{\;\;A}_Cn_An^C g_{\mu\nu}
-{}^{(5)}R^A_{\;\;BCD}n_An^Cg_{\mu}^Bg_{\nu}^D\nonumber\\
&& - \frac{1}{2}{}^{(5)}R^A_{\;\;BCD}n^Bn^Cn^Dn_A g_{\mu\nu} + ({\rm Tr}\; K) K_{\mu\nu} - K^C _{ \nu}K_{\mu C}
 - \frac{1}{2}g_{\mu\nu}(({\rm Tr}\; K)^2 - ({\rm Tr}\; K^2)).
\end{eqnarray}
Using Einstein equations \cite{mis}, it is possible to write
\begin{equation}
{}^{(5)}G_{AB} = - \frac{1}{2}{\Lambda}_5{}^{\; (5)}g_{AB} + \kappa_5^2{}^{(5)}T_{AB},
\end{equation}
\noindent
where $\Lambda_5$ denotes  the cosmological constant in the bulk, which
prevents gravity from leaking into the extra dimension at low energies. The field equations on the brane can be written as
\begin{eqnarray}\label{field}
G_{\mu\nu}  &=& -\frac{1}{2}{\Lambda}_5g_{\mu\nu} + \kappa_5^2{}^{(5)}T_{\mu\nu} +
{}^{(5)}R_{BD}n^Bn^Dg_{\mu\nu} +
\frac{1}{2}{}^{(5)}R^{\;\;A}_Cn_An^C g_{\mu\nu}\nonumber\\
&&-{}^{(5)}R^A_{\;\;BCD}n_An^Cg_{\;\;\mu}^Bg_{\;\;\nu}^D
 - \frac{1}{2}{}^{(5)}R^A_{\;\;BCD}n^Bn^Cn^Dn_A g_{\mu\nu} + ({\rm Tr}\; K) K_{\mu\nu} - K^C _{ \nu}K_{\mu C}\nonumber\\
&& - \frac{1}{2}g_{\mu\nu}(({\rm Tr}\; K)^2 - ({\rm Tr}\; K^2)).
\end{eqnarray}
\noindent where $\kappa_5 = 8\pi G_5$. The Riemann tensor can be expressed as
a combination of the Weyl tensor, the Ricci tensor and the Ricci scalar, as
$^{(5)}R^A_{\;\;BCD} = {}^{(5)}C^A_{\;\;BCD} + {}^{(5)}D^A_{\;\;BCD}$, where
$^{(5)}C^A_{\;\;BCD}$ denotes the Weyl tensor and $^{(5)}D^A_{\;\;BCD}$ can be written as \cite{Shiromizu}
\begin{eqnarray} \label{weyl}
{}^{(5)}D^A_{\;\;BCD} &=& \frac{2}{3}\left({}^{(5)}g^A_{\;\;C}{}^{(5)}R_{DB} - {}^{(5)}g^A_{\;\;D}
{}^{(5)}R_{CB} - {}^{(5)}g_{BC}{}^{(5)}R^A_{\;\;D} + {}^{(5)}g_{BD}{}^{(5)}R^A_{\;\;C}\right)
 \nonumber\\
&&- \frac{1}{6}\left({}^{(5)}g^A_{\;\;C}{}^{(5)}
g_{BD} - {}^{(5)}g^A_{\;\;D}{}^{(5)}g_{BC}\right){}^{(5)}R.
\end{eqnarray}

\section{Energy conservation and junction conditions}

From eqs.(\ref{field}) and (\ref{weyl}) it follows that
\begin{eqnarray}\label{einstein}
G_{\mu\nu}  &=& -\frac{1}{2}{\Lambda}_5g_{\mu\nu} +  \frac{2}{3}\kappa_5^2\left[^{(5)}T_{AB}g_{\;\;\mu}^{A}g_{\;\;\nu}^{B} + \left(^{(5)}T_{AB}n^An^B - \frac{1}{4}({\rm Tr}\; ^{\; (5)}T)\right)g_{\mu\nu}\right]\nonumber\\
&&+ ({\rm Tr}\; K) K_{\mu\nu} - K^C _{ \nu}K_{\mu C} + \frac{1}{2}g_{\mu\nu}[({\rm Tr}\; K)^2 - ({\rm Tr}\; K^2)] - E_{\mu\nu}.
\end{eqnarray}
\noindent
The term $E_{\mu\nu}$ is the projection of the bulk Weyl tensor on the brane:
\begin{equation}
E_{\mu\nu} = {}^{(5)}C_{ACBD}n^Cn^Dg_{\;\;\mu}^{A}g_{\;\;\nu}^{B}.
\end{equation}

Now, the curvature factors are to be expressed in terms of
momentum-energy tensors. This can be done using two conditions: the Israel-Darmois junction conditions and the $\mathbb{Z}_2$-symmetry.
This last one physically asserts that, when one approaches the brane from one side and go through it, she emerges into a bulk that looks the same,
but with the normal vector direction reversed,
$n \mapsto -  n$, meaning a mirror condition. In more formal terms, taking another basis $\{v_A\}$ of the bulk,  it is well known
that the exterior product among elements of each basis $\{e_A\}$ and $\{v_A\}$ are related by
$v_0\wedge v_1\wedge\cdots\wedge v_4 = {\rm det}(v_A^{\;\;B})\;e_0\wedge e_1\wedge \cdots\wedge e_4$, where $v_A = v_A^{\;\;B}e_B$.  The number
${\rm det}(v_A^{\;\;B})$ is positive [negative]
if the bases $\{e_A\}$ and $\{v_A\}$ have the same [opposite] orientation, where the orientation in  the bulk
is defined as a choice in $\mathbb{Z}_2$ of equivalence classes of bases  in the bulk
 (or also in the brane).
Any tangent vector space  has only two orientation possibilities, depending of the signal of  ${\rm det}(v_A^{\;\;B})$, and
any homomorphism given by $e_A\mapsto F\,e_A$, satisfying ${\rm det}(F_A^{\;\;B}) > 0$,
does not change the tangent space orientation. Then the mirror condition means the orientation over $T_xM$ is changed. It is worth to emphasize
that such mirror transformations are trivially achieved in bulks described by non-orientable manifolds, like the M\"obius strip, for instance.
In this case it is possible to achieve mirror transformations without puncturing the surface, by means of the parallel transport of a normal
vector field, over a closed path. In this sense, after parallel transporting the vector, normal to the surface, over a closed path,
it points in the opposite direction and, despite
reversing the normal vector direction, the related observer is on the another side of the brane.
The formalism of brane world scenarios on non-orientable manifolds are beyond the scope of the present paper.

The Israel-Darmois conditions \cite{is, dar, lanc, sen, ko} are obtained assuming that in the bulk field equations,
\begin{equation}
^{(5)}G_{AB} = - \frac{1}{2}{\Lambda}_5{}^{\; (5)}g_{AB} + \kappa_5^2{}^{(5)}T_{AB},
\end{equation}\noindent  the momentum-energy tensor $^{(5)}T_{AB}$ is a sum of a bulk intrinsic momentum-energy tensor
 and a brane momentum-energy tensor integrated in a brane neighbour region
(from $y = - \epsilon$ to $y = + \epsilon$, where $y$ is the coordinate representing the bulk from the brane). The Israel-Darmois conditions are given by
\begin{equation}
g^+_{\mu\nu} - g^-_{\mu\nu} = 0,\qquad\qquad K^+_{\mu\nu} - K^-_{\mu\nu} = -\kappa^2_5\left(T_{\mu\nu}^{\rm brane} - \frac{1}{3}T^{\rm brane}g_{\mu\nu}\right),
\end{equation}
\noindent
where $T^{\rm brane} = ({\rm Tr}\; T^{\rm brane}_{\mu\nu})$. The term $T_{\mu\nu}^{\rm brane}$
is the description of the momentum-energy tensor of the brane by an \emph{inhabitant} of the bulk. In order to write $T_{\mu\nu}^{\rm brane}$ using entities
realized by an inhabitant in the brane, we use
\begin{equation}
T_{\mu\nu}^{\rm brane} = T_{\mu\nu} - {\lambda}g_{\mu\nu},
\end{equation}
\noindent
where $\lambda$ is defined as the tension on the brane.
The second assumption is the $\mathbb{Z}_2$-symmetry, resulting in the conditions
 $K^+_{\mu\nu} = - K^-_{\mu\nu} = K_{\mu\nu}$. Substituting these expressions  in the Israel-Darmois conditions it follows that
\begin{equation}
K_{\mu\nu} = -\frac{1}{2}\kappa^2_5\left(T_{\mu\nu} + \frac{1}{3}(\lambda - T)g_{\mu\nu}\right).
\end{equation}
Another step is to use the results above in order to develop the extrinsic curvatures terms in eq.(\ref{einstein}):
\begin{equation}\label{adriana}
({\rm Tr}\; K) K_{\mu\nu} - K^C _{ \nu}K_{\mu C} = \frac{1}{4}\kappa^4_5\left(TT_{\mu\nu} - T^C _{ \nu}T_{\mu C}\right),
\end{equation}\noindent obviously implying that
\begin{equation}\label{tamara}
({\rm Tr}\; K)^2 - ({\rm Tr}\; K^2) = \frac{1}{4}\kappa^4_5\left(({\rm Tr}\; T)^2 - ({\rm Tr}\; T^2)\right).
\end{equation} From eqs.(\ref{adriana}) and (\ref{tamara}) it follows that
\begin{eqnarray}\label{curvaturas}
({\rm Tr}\; K) K_{\mu\nu} - K^C _{ \nu}K_{\mu C} + \frac{1}{2}g_{\mu\nu}(({\rm Tr}\; K)^2 - ({\rm Tr}\; K^2))= \frac{1}{4}\kappa_5^4\left[TT_{\mu\nu} - T^C _{ \nu}T_{\mu C} + \frac{1}{2}g_{\mu\nu}(({\rm Tr}\; T)^2 - ({\rm Tr}\; T^2))\right].&&
\end{eqnarray}
\noindent
Substituting eq.(\ref{curvaturas}) in eq.(\ref{einstein}), we obtain the following equation for the fields in the brane:
\begin{eqnarray}
G_{\mu\nu}  &=& -\frac{1}{2}{\Lambda}_5g_{\mu\nu} +  \frac{2}{3}\kappa_5^2\left[^{\; (5)}T_{AB}g_{\;\mu}^{A}g_{\;\nu}^{B} + \left(^{\; (5)}T_{AB}n^An^B - \frac{1}{4}({\rm Tr}\; ^{\; (5)}T)\right)g_{\mu\nu}\right]\nonumber\\
&& + \frac{1}{4}\kappa_5^4\left[TT_{\mu\nu} - T^C _{ \nu}T_{\mu C} + \frac{1}{2}g_{\mu\nu}(({\rm Tr}\; T)^2 - ({\rm Tr}\; T^2))\right] - E_{\mu\nu}.
\end{eqnarray}
This equation can be further simplified if the momentum-energy conservation law is valid on the brane, i.e.,
\begin{equation}
T_{\mu\nu}^{\; \; \; ;\nu} = 0.
\end{equation}
\noindent
An expression for $T_{\mu\nu}^{\; \; \;  ;\nu}$ in terms of $^{(5)}T_{AB}$ can be found by combining  the following equations, respectively
describing the field equations in the bulk, the Israel-Darmois junction conditions and the Gauss-Codazzi equations:
\begin{equation}\label{fields}
^{(5)}G_{AB} = - \Lambda_5 ^{\; (5)}g_{AB} + \kappa^2_5 {}^{\; (5)}T_{AB},
\end{equation}
\begin{equation}\label{israel}
K_{\mu\nu} = -\frac{1}{2}\kappa^2_5\left[T_{\mu\nu} + \frac{1}{3}(\lambda - T)g_{\mu\nu}\right],
\end{equation}
\noindent
\begin{equation}\label{codazzi}
K^{\;\; B}_{A\; \; ;B} - K_{;A} = {}^{(5)}R_{BC}g_{\;\;A}^{B}n^C.
\end{equation}
\noindent
Performing the covariant derivative of the curvatures and using eq.(\ref{israel}) it follows that
\begin{equation}\label{facial}
K_{\mu\nu}^{ \; \; ; \nu} - K^{;\nu} = - \frac{1}{2}\kappa_5^2T_{\mu\nu}^{\; \;  ; \nu},
\end{equation}
\noindent
and using eq.(\ref{codazzi}) it follows that
\begin{equation}
^{\; (5)}R_{AB}g^A_{\;\; \nu}n^B = K^{\; A}_{\nu\; \; ; A} - K_{;\nu}.
\end{equation}
\noindent
But from eq.(\ref{fields}), the expression
\begin{equation}
^{(5)}R_{AB} = - \Lambda_5 ^{(5)}g_{AB} + \kappa^2_5 {}^{(5)}T_{AB} + \frac{1}{2} {}^{(5)}g_{AB}{}^{(5)}R
\end{equation}
\noindent
holds, which implies that
\begin{equation}\label{desenvolvimento}
K^{\; A}_{\nu\; \;  ; A} - K_{;\nu} = \left(- \Lambda_5 ^{\; (5)}g_{AB} + \kappa_5 ^{2\; (5)}T_{AB} + \frac{1}{2}{}^{(5)}g_{AB}{}^{(5)}R\right)g_{A\nu}n_B.
\end{equation}
\noindent
Now, introducing eq.(\ref{neo}) in eq.(\ref{desenvolvimento}) and attempting for the fact that the projections of the Ricci tensors in the $n^A$ direction are zero,
we get from eq.(\ref{facial})
\begin{equation}
T^{\; A}_{\mu \; \; ;A} = - 2^{\; (5)}T_{AB}g_{\mu}^{\; A}n^B.
\end{equation}
\noindent
Assuming that $T^{\; A}_{\mu \; \; ;A} = 0$ in the brane, it is immediate that $^{(5)}T_{AB} = 0$.
This means the bulk is in complete vacuum and the particles are in fact on the brane. Now, the field equations reduce to
\begin{eqnarray}\label{123}
G_{\mu\nu}  &=& -\frac{1}{2}{\Lambda}_5g_{\mu\nu}
+ \frac{1}{4}\kappa_5^4\left[TT_{\mu\nu} - T^C _{ \nu}T_{\mu C} + \frac{1}{2}g_{\mu\nu}(({\rm Tr}\; T)^2 - ({\rm Tr}\; T^2))\right] - E_{\mu\nu},
\end{eqnarray}
\noindent
showing the contribution of the bulk on the brane is only due to the Weyl tensor.

\section{Black holes on the brane}
The Bianchi identities, applied to the Einstein field equations found in the last section, are expressed as
$G_{\mu\nu}^{\; \; \; ;\nu} = 0$, which implies that $E_{\mu\nu}^{\; \; \; ;\nu} - S_{\mu\nu}^{\; \; \; ;\nu}= 0,$
where
\begin{equation}
S_{\mu\nu} := \frac{1}{4}\kappa_5^4\left[TT_{\mu\nu} - T^C _{ \nu}T_{\mu C} + \frac{1}{2}g_{\mu\nu}(({\rm Tr}\; T)^2 - ({\rm Tr}\; T^2))\right].
\end{equation}
A vacuum on the brane, where $T_{\mu\nu} = 0$ outside a black hole, implies that
\begin{equation}\label{21}
E_{\mu\nu}^{\; \; \; ;\nu} = 0.
\end{equation}
\noindent
Eqs.(\ref{21}) are referred to the non-local conservation equations. Other useful equations for the black hole case are
\begin{equation}\label{ricci2}
G_{\mu\nu} = - \frac{1}{2}\Lambda_ 5g_{\mu\nu} - E_{\mu\nu}, \qquad\qquad R = R^{\mu}_{\;\; \mu} = 0 = E^{\mu}_{\;\; \mu}.
\end{equation}
\noindent
The Weyl term $E_{\mu\nu}$  carries an imprint of high-energy effects sourcing KK modes. It means that highly energetic
stars and the process of gravitational collapse, and naturally  black holes, lead to deviations from the $4D$ general
relativity problem. This occurs basically because the gravitational collapse unavoidably produces energies high enough to make
these corrections significant. From the brane-observer viewpoint, the KK corrections in $E_{\mu\nu}$ are nonlocal, since they
incorporate $5D$ gravity wave modes. These nonlocal corrections cannot be determined purely from data on the brane \cite{Maartens}.
In the perturbative analysis of Randall-Sundrum (RS) brane, KK modes generate a correction in the gravitational potential $V(r) = GM/r$, which is
given by \cite{Randall}
\begin{equation}\label{potential}
V(r) \approx \frac{GM}{r}\left(1 + \frac{2l^2}{3r^2}\right).
\end{equation}
\noindent
The KK modes that generate this correction are responsible for a nonzero $E_{\mu\nu}$. This term carries the modification to the weak-field field equations, as we have
already seen.
The RS metric is in general expressed as
\begin{equation}
^{(5)}ds^2 = e^{-2k|y|}g_{\mu\nu}dx^{\mu}dx^{\nu} + dy^2,\qquad 0\leq y\leq \pi r_c
\end{equation}
\noindent
where $k^2 = 3/(2l^2)$, $l$ denotes the bulk curvature radius, $k$ is a scale of order the Planck scale,
 $r_c$ is the compactification radius and the term $e^{-2|y|/l}$ is called the \emph{the warp factor} \cite{Randall}, which
reflects the confinement role of the bulk cosmological constant $\Lambda_5$, preventing gravity from leaking
into the extra dimension at low energies.

Concerning the anti-de Sitter (AdS$_5$) bulk, the cosmological constant can be written as $\Lambda_5 = -6/l^2$,
where $l$ is the curvature radius of AdS$_5$.
The brane is localized
at $y = 0$, where the metric recovers the usual aspect. Therefore, a particular manner to express the vacuum field equations in the brane given by eq.(\ref{ricci2}) is
\begin{equation}
E_{\mu\nu} = - R_{\mu\nu},
\end{equation}
\noindent
where the bulk cosmological constant is incorporated to the warp factor in the metric.
One can use a Taylor expansion in order to probe properties of a static black hole on the brane \cite{Dadhich}, and for a vacuum brane metric,
\begin{eqnarray}\label{metrica}
g_{\mu\nu}(x,y) &=& g_{\mu\nu}(x,0) - E_{\mu\nu}(x,0)y^2 - \frac{2}{l}E_{\mu\nu}(x,0)y^3
+\frac{1}{12}\left({\Box}E_{\mu\nu} - \frac{32}{l^2}E_{\mu\nu} + 2R_{\mu\alpha\nu\beta}E^{\alpha\beta} + 6E_{\mu}^{\; \alpha}E_{\alpha\nu}\right)_{y=0}{\; \; }y^4 + \cdots
\nonumber\end{eqnarray}
\noindent where $\Box$ denotes the usual d'Alembertian.
It shows in particular that the propagating effect of $5D$ gravity arises only at the fourth order of the expansion. For a static spherical metric on the brane
given by \begin{equation}\label{124}
g_{\mu\nu}dx^{\mu}dx^{\nu} = - F(r)dt^2 + \frac{dr^2}{H(r)} + r^2d\Omega^2,
\end{equation}
\noindent
 the projected Weyl term on the brane is given by the expressions
\begin{equation}
E_{00} = \frac{F}{r}\left(H' - \frac{1 - H}{r}\right),\qquad
E_{rr} = -\frac{1}{rH}\left(\frac{F'}{F} - \frac{1 - H}{r}\right),
\qquad E_{\theta\theta} = -1 + H +\frac{r}{2}H\left(\frac{F'}{F} + \frac{H'}{H}\right).
\end{equation}
\noindent Note that in eq.(\ref{124}) the metric is led to the Schwarzschild one, if $F(r)$ equals $H(r)$.
The exact determination of these radial functions remains an open problem in black hole theory on the brane \cite{Maartens}.

These components allow one to evaluate the metric coefficients in eq.(\ref{metrica}). The area of the
$5D$ horizon is determined by $g_{\theta\theta}$. Defining $\psi(r)$ as the deviation from a Schwarzschild form $H$, i.e.,
\begin{equation}\label{h}
H(r) = 1 - \frac{2M}{r} + \psi(r),
\end{equation}
\noindent
where $M$ is constant, yields
\begin{equation}\label{gtheta}
g_{\theta\theta}(r,y) = r^2  + \psi'\left(1 + \frac{2}{l}y\right)y^2 + \frac{1}{6r^2}\left[\psi' + \frac{1}{2}(1 + \psi')(r\psi' - \psi)'\right]y^4 + \cdots
\end{equation}
\noindent
It shows how $\psi$ and its derivatives determine the change in the area of the horizon along the extra dimension.
 For a large black hole, with horizon scale $r \gg l$, we have from eq.(\ref{potential}) that
\begin{equation}\label{psi}
\psi \approx -\frac{4Ml^2}{3r^3},
\end{equation}
\noindent
that shall be used to estimate the AdS$_5$ bulk curvature radius due to
 observational luminosities of quasars, that is the same as to the supermassive black holes accreting gas.
\section{Luminosity in quasars and AdS curvature radius}
The observation of quasars (QSOs) in $X$-ray band can constrain the measure of the AdS$_5$ bulk curvature radius $l$, and indicate
 how the bulk is curled, from its geometrical and topological features.
QSOs are astrophysical objects that can be found at large astronomical distances (redshifts $z > 1$).
For \emph{gedanken} experiment involving a static black hole being accreted, in a simple model, the accretion eficiency $\eta$ is given by
\begin{equation}\label{eta}
\eta = \frac{M}{6R_{{\rm Sbrane}}},
\end{equation}
\noindent
where $R_{{\rm Sbrane}}$ is the Schwarzschild radius corrected for the case of brane-world. The luminosity $L$ due to accretion in a black hole, that generates a quasar,
 is given by
\begin{equation}
L = \eta \dot{M},
\end{equation}
\noindent
where $\dot{M}$ denotes the accretion rate and depends on some specific model of accretion.
In order to estimate $R_{{\rm Sbrane}}$, fix $H(r) = 0$ in  eq.(\ref{h}), resulting in
\begin{equation}
1 - \frac{2M}{R_{{\rm Sbrane}}} - \frac{4Ml^2}{3R_{{\rm Sbrane}}^3} = 0.
\end{equation}
\noindent
The above solution for $R_{{\rm Sbrane}}$ is relatively straightforward to be found in terms of the curvature radius $l$.
It is then possible to find an expression for the luminosity $L$  in terms of the radius of curvature
\begin{equation}
L(l) = \eta(l) \dot{M}.
\end{equation}
\noindent
Having observational values for the luminosity $L$, it is  possible to estimate a value for $l$, given a black hole accretion model.
As usual,  denoting $M_{\odot} = 10^{33}$ g  the mass of the sun, for a tipical supermassive black hole of $10^9 M_{\odot}$ in a massive quasar,
a spherical accretion model as the Bondi-Hoyle model \cite{Bondi} gives $\dot{M} \sim 800 M_{\odot}\;{\rm yr}^{-1}.$
Supposing the quasar radiates in Eddington limit, given by (see, e.g., \cite{shapiro})
\begin{equation}
L(l) = L_{Edd} = 1.26 \times 10^{45}\left(\frac{M}{10^7 M_{\odot}}\right)\; {\rm erg\,  s^{-1}}
\end{equation}
\noindent
for a quasar with a supermassive black hole of $10^9 M_{\odot}$, the luminosity is given by $L \sim 10^{47}\, {\rm erg\, s^{-1}}$.
The accretion efficiency, in this case, is given by
$\eta(l) \sim 10^{-2}$ or, from eq.(\ref{eta}),
$R_{{\rm Sbrane}}(l) \sim 10M.
$ Estimating $R_{{\rm Sbrane}}(l)$ in terms of $l$ is equivalent to estimate the curvature radius of the AdS$_5$ bulk.
Substituting the result above in eq.(\ref{h}) gives $\psi \approx -0.8$, with respect to a SMBH with mass of magnitude $10^9 M_\odot$.
In the first approximation given by eq.(\ref{psi}) the AdS$_5$ curvature radius is $l \sim 25M$.
It accords to the WMAP observational results \cite{wmap}, since it agrees with the possibility of an approximately flat
universe at large scales.

Now we are able to determine the order of the extra dimension compactification scale $k$, since it is closely related to
the AdS$_5$ bulk curvature radius \cite{Randall} by the expression $k^2 = 3/(2l^2)$, leading to  $k \approx 10^{-2}/M$, where
it must be emphasized $M$ denotes a SMBH mass.
This can be useful for the future accelerators experiments in order to impose limits on the investigation range of the extra dimension
compactification radius, among other branches of applications.

\section{Concluding Remarks}

The outcomes presented in the last Sections points basically for two possible conclusions. On the one hand, the AdS$_5$
bulk curvature radius, and consequently the extra dimension compactification radius, is determinated from quasar luminosity observations.
On the other hand, the approximation given by eq.(\ref{psi}), proposed by a lot of authors, is not a so good solution in order to describe a Schwarzschild black hole on the brane.
In fact, for the approximation adopted, $g_{\theta\theta}$ in eq.(\ref{gtheta})  decreases as one moves off the brane. This is consistent with a
pancake-like shape of the horizon. However, the correct horizon shape is the tubular one, in Gaussian normal coordinates (see, e.g.,  \cite{Gian}).  Thus
a simple brane-based approach, while giving useful insights, does not lead to a very realistic black hole solution. Indeed, up to this moment, there is no known
 solution representing a realistic black hole localized on the brane, which is stable and does not present any naked singularity \cite{Maartens}.
This remains a key open question on nonlinear brane-world gravity. Some applications in astrophysics \cite{darocha3} and particle physics \cite{coimbra}, involving extensions of the present paper, are presented \cite{darocha3,coimbra}.

However, the points discussed in the present paper sheds new light on the theory of supermassive black holes in brane-world scenarios,
since it exhibits a constraint involving quasars luminosity and the AdS$_5$ curvature radius, in order to find a
suitable black hole solution. If the brane-world models are
to be correct,  the quasars luminosity does contain a smoke gun signature to estimate the right correction of the term $\psi$ in eq.(\ref{h}),
 and the correct fixed value to the AdS$_5$ curvature radius $l$.

One interesting remark is that the AdS$_5$ influence on the brane is totally due to the Weyl tensor, related to the bulk geometry. From the
 geometric viewpoint, it carries intrinsic modifications in the black hole accretion mechanism.
 Some alternative directions to approach the present formalism can be pointed out, trying to encompass the questions related to this article.
The first one arises from the fact that, since after Witten proved
general relativity is a renormalizable quantum system in (1+2) dimensions \cite{Witten},
a lot of interesting motivations to investigate the AdS$_5$ spacetime have been arising, in the light of the generalization
of the gravity gauge theory in (1+2) dimensions to higher ones \cite{darocha}. The first attempt was to enlarge the Poincar\'e group of symmetries,
 supposing an AdS group symmetry \cite{tron}, which contains the Poincar\'e group. Also, the AdS/CFT correspondence
 asserts that a  maximal supersymmetric Yang-Mills theory in 4-dimensional Minkowski-spacetime is equivalent to a type IIB closed
superstring theory \cite{gr, hoker, mald1, mald2, mald3, mald4, mald5, mald15}.
The 10-dimensional arena for the type IIB superstring theory is then described by the product manifold
$S^5\times$ AdS$_5$, and so our future investigations are directed towards brane-world scenarios and AdS$_5$ spacetime and their
possible relationships with supersymmetric theories.

Other two possible directions to find black hole solutions in a brane-world scenario involve respectively the cohomological
and geometrical aspects of the brane and the bulk.
On the one hand, nowadays the search for a brane-world scenario is based on an orientable manifold describing the bulk. The $\mathbb{Z}_2$-symmetry in the brane,
known as the mirror symmetry, describes a change in the orientation when one passes through the brane, from one side to the other side
of the brane. It can be interpreted by a non-homomorphic `puncture' map on the brane, which can be led to a homomorphic map, if one supposes that the brane
is \emph{non-orientable} \cite{darocha}. For instance, considering the non-orientable M\"obius strip,
the parallel transport of a normal vector on a closed path makes the vector reverse the sign after
returning to the initial point over the strip. Up to our knowledge, there is no investigation about branes in non-orientable manifolds, which certainly can
bring some novelty in this open research field.
On the another hand, the Weyl tensor
 has been more formally investigated by its increasing importance concerning symplectic
geometry and twistor fibrations, which are closely related to open problems in quantum gravity \cite{giu, gu1}.
It is known that  the set of all compatible complex structures on the
tangent spaces of the bulk $M$ has twistor structure, by considering a canonical almost complex structure induced by a symplectic connection on $M$. The
integrability equation of the almost complex structure appears in terms of the curvature and it agrees with the vanishing
 of the Weyl curvature tensor, related to conformal flatness. It proposes a close relation between the investigations on symplectic connections and that of twistor spaces \cite{beng},
and the recent developments in string theory \cite{b,b1}.
More comments concerning such structures are beyond the scope of the present paper.

\section{Acknowledgements}
The authors are grateful to Prof. Paul K. Townsend for his comments and suggestions, to Prof.
Roy Maartens for his patience and clearing up expositions concerning branes, to Julio Hoff M. da Silva for pointing out some missing points,
and specially to Dr. Ricardo A. Mosna for his careful reading of the manuscript
and enlightening  suggestions.

\section*{Appendix: some useful properties of the Weyl tensor}

 The curvature tensor can be decomposed into the part which depends on the Ricci curvature, and the Weyl tensor.
Considering a manifold of dimension $n$, when $n > 3$ the second part can be non-zero.
In a conformal change of the spacetime metric $g$, the Weyl tensor is invariant, and it is also called conformal tensor.
A necessary condition for a Riemannian manifold to be conformally flat is that the Weyl tensor vanish.
 Moreover, the Weyl tensor is identically null if and only if
the metric is locally conformal to the standard Euclidean metric.

The Weyl curvature tensor is defined as being the traceless component of the Riemann curvature tensor.
 In other words, it is a tensor that has the same symmetries as the Riemann curvature tensor with the extra condition that its Ricci curvature must vanish \cite{mis}.

In dimensions two and three the Weyl curvature tensor vanishes identically.
The Weyl tensor can be obtained from the full curvature tensor by subtracting out various traces.
This is most easily done by writing the Riemann tensor as a (0,4) valent tensor, by contracting with the metric. The (0,4) valent Weyl tensor is then written as
$$
W = R - \frac{1}{n-2}(Ric - \frac{s}{n}g)\circ g - \frac{s}{2n(n-1)}g\circ g
$$\noindent
where $Ric$ denotes the Ricci tensor, $R$ denotes the scalar curvature, and $g\circ h$  denotes the
Kulkarni-Nomizu product of two arbitrary metrics $g$ and $h$:
$$(g\circ h) (u,v,w,z):= g(u,w)h(v,z) + h(u,w)g(v,z) - g(u,z)h(v,w) - g(v,w)h(u,z).$$

The ordinary (1,3) valent Weyl tensor is then given by contracting the above with the inverse of the metric.
The Ricci tensor describes the local gravitational field of the nearby matter. The long-range gravitational field,
namely gravitational waves and tidal forces, propagates through the Weyl conformal curvature tensor. It is well known that spacetime, at least locally, can be split
in spacelike surfaces plus time, if a timelike vector field $n$,  tangent to the worldline of a given observer, is chosen.
 The splitting of the gravitational field into its local and nonlocal components is shown in the following decomposition of the Riemann tensor
\begin{equation}
 R_{abcd} = C_{abcd} + \frac{1}{2} (g_{ac} R_{bd} + g_{bd} R_{ac} -  g_{bc} R_{ad} -   g_{ad} R_{bc}) + \frac{1}{6} R ( g_{ac} g_{bd} -   g_{ad} g_{bc}),
\end{equation}\noindent where $C_{abcd}$ denotes the components of the  Weyl tensor, which presents all the symmetries of the Riemann tensor and also $C^c_{\;\;acb} = 0$.
Relative to the fundamental observers, the Weyl tensor decomposes further into its irreducible parts according to
\begin{equation}
C_{abcd} = (g_{abqp} g_{cdsr} - \eta_{abqp}\eta_{cdsr}) n_q n_s E^{pr} +  (\eta_{abqp} g_{cdsr} + g_{abqp}\eta_{cdsr}) u_q u_s H^{pr}
\end{equation}\noindent where $g_{abcd} := g_{ac}g_{bd} - g_{ad} g_{bc}$ \cite{scheilamelo, tsagas}. The symmetric and trace-free components $E_{ab}$  and $H_{ab}$  are known as
 the electric and magnetic Weyl components and they are given by $E_{ab} = C_{acbd} u^c u^d$ and $H_{ab} = \frac{1}{2}\epsilon_a^{cd} C_{cdbe} u^e$,
with $E_{ab} u^b = 0 = H_{ab} u^b$. The electric part of the Weyl tensor is associated with the tidal field.
The magnetic component is associated with gravitational waves \cite{anegrini}. Of course, both tensors are required if gravitational waves are to exist.
 The Weyl tensor represents the part of the curvature that is not determined locally by matter.
The dynamics of the Weyl field is not entirely arbitrary because the Riemann tensor satisfies the Bianchi identities, that
when contracted take the form \cite{scheilamelo}
\begin{equation}
\nabla\label{1} C_{abcd} =  \nabla_{[b} R_{a]c} + \frac{1}{6} g_{c[b}\nabla_{a]} R.
\end{equation}\noindent Then, field equations for the Weyl tensor are simulated by Bianchi identitites,
determining the part of the spacetime curvature that depends on the matter distribution at other points \cite{scheilamelo}.
 Eq.(\ref{1}) splits into a set of two propagation and two constraint equations, which gives the evolution of the electric and magnetic Weyl components
\cite{1}.

\end{document}